\documentstyle[preprint,aps]{revtex}

\begin{document}
\draft
\preprint{}
\title{Analytical model for a crossover between uncorrelated and fractal
behaviour of a self-repulsive chain}
\author {A.N. Rubtsov}
\address{Physics
Department, Moscow State University, 119899 Moscow, Russia}
\date{\today}
\maketitle
\begin{abstract}
The thermodynamics of a long self-repulsive chain is studied.
In $D<4$ dimensions it shows
two distinctly different regimes, corresponding to weak and strong
correlations in the system.
A simple microscopic analytical model is presented
which successfully describes both the regimes.
The self-consistent scheme is used, in which
the center of mass of a chain
is fixed explicitly. This allows to take correlations
into account in an indirect manner.
\end{abstract}

\pacs {05.20.Gg, 05.40.Fb}

\narrowtext

Problem of a crossover between the strong-correlation regime and
weak correlations is of a general interest for the solid-state
physics. A textbook example is here the theory of phase transitions \cite{LL5}.
Not too close to the transition point Landau theory is valid,
yielding the dimension-independent Landau set of critical indices.
On the other hand, correlations in a critical region in a very vicinity
of the transition result
in different (dimension-dependent) values of critical indices,
as it can be described by the renormalisation-group analysis \cite{PP,BC}.
While the two regimes are well understood in itself,
no general microscopic analytical formula describes them simultaneously.

In this paper we draw a formula of this kind, which describes
the behaviour of a simpler system, namely the self-repulsive chain.
The peculiarity of this
system is that it does not show
a phase transition, but the two regimes mentioned above are still present.
We believe that an absence of a spontaneous symmetry break makes
things simpler, while the physics of a crossover
remains somehow similar to more complicated problems like the description
of the phase transition; for instance the problem of a
self-interacting chain is closely related to the theory of percolation \cite{Sokolov}.

Let us first define the formal (mathematical) problem.
We consider paths of a random walker on a cubic
lattice in $D$ dimensions, which returns to the starting point
after $L$ steps (see Figure 1 for the illustration);
we suppose that $L\gg 1$. These paths are closed polygons (maybe,
self-crossing); positions of their vertexes $x_i$ fulfill the condition
\begin{equation}  \label{E1}
  ||x_i-x_{i+1}||=1;~~~x_0=x_L.
\end{equation}
We define the number of self-crosses for a given path $X=\{x_0, x_1, ..., x_{L-1}\}$
by the formula
\begin{equation}
N(X)=\frac{1}{2} \sum_{i\neq j} \delta(x_i;x_j),
\end{equation}
where $\delta$ is a Kronecker symbol, and introduce
a statistical average as follows:
\begin{eqnarray} \label{avs}
  <f>&=&Z^{-1} \sum_{X} f(X) e^{-\beta N(X)};\\
  \nonumber
  Z&=&{\sum_{X} e^{-\beta N(X)}}.
\end{eqnarray}
Here sums are taken over all paths $X$, $f$ is a function of $X$,
and $\beta$ is a small positive
parameter. In particular, we shall monitor the dispersion $<(x-u)^2>$,
where $u=L^{-1}\sum_i x_i$ is the position of the center of mass of a chain.

Physically, the above equations describe the statistics of a loop-chain with a potential
energy equal to the number of its self-crosses $N$,
at the inverse temperature $\beta$.

Since $\beta \ll 1$, for a sufficiently small $L$ the statistical weight $e^{-\beta N}$
is always close to 1, {\it i.e.} correlations
in the system are negligible.
Therefore we deal with a standard random-walk
problem; for this case $<(x-u)^2> \propto L$.
We shall refer
to this case as an uncorrelated regime.

Let us establish a parameter which determines the magnitude of
fluctuations. Since the typical volume occupied by the chain  is
proportional to $L^{D/2}$ in the uncorrelated regime,
the  average number of its self-crosses  can be estimated
as $N\propto L^{2-D/2}$. Therefore the dimensionless
length $\lambda=\beta^{2/(4-D)} L$ can be introduced; the
uncorrelated regime occurs at $\beta N \ll 1$, that corresponds to
$\lambda \ll 1$.

In the opposite case of a large $\lambda$, correlations in the chain
affect its statistics strongly.
The problem is similar to the well-studied case of the statistics
of a self-avoiding polymer molecule \cite{RMP,Sokolov}.
The scaling laws are dimension-dependent
and can be calculated by the renormalisation-group technique.
For this "fractal" limit $<(x-u)^2> \propto L^\phi$ with
$\phi=2$ in 1D and takes non-integer values in 2D and 3D, which are
respectively close to
4/3 and 6/5.  In
$D\geq4$ the fractal regime is virtually absent, as
$\phi$ takes the "uncorrelated" value 1 (there are logarithmic corrections
in $D=4$).

As it was declared above, the present paper is aimed to
establish a formula that describes (at least qualitatively)
both the uncorrelated regime and the fractal regime.
The outline is as follows.
First, the continuous analog (\ref{cont}) of the chain under
study is introduced. Then, we switch to a kind of the self-consistent
potential approximation (\ref{repl}).
It yields simple formulae (\ref{Smin}-\ref{eq}), that show
the desired asymptotes for $<(x-u)^2>$. The result is then
compared with numerical calculations.

Let us construct a continuous analog. One can neglect the short-range correlations,
as the critical behaviour in $D<4$ is due to the long-range ones \cite{PP}.
Let $\tau \gg 1$ be a "physically infinitesimal" length, on which
correlations can be neglected; accordingly to the above estimations
it should be $\tau\ll \beta^{-2/(4-D)}$.  Since
$\beta$ is small, in $D<4$ these two inequalities can be
fulfilled simultaneously.
Consider the paths passing through some $x_i$. Then the distribution
function of $x$ at $i+\tau$ is given by
\begin{equation} \label{Gauss}
p_{i+\tau}(x;x_i)=p_0 \exp\left(-\frac{D (x-x_i)^2}{2 \tau}\right),
\end{equation}
where $p_0(\tau)$ is a normalisation factor.
The number of paths passing {\it simultaneously} through given $x_i$ and
$x_{i+\tau}$, is obviously proportional to $p_{i+\tau}(x_{i+\tau};x_i)$.

Introduce also the "partial" average number of self-crosses
$n_{ij}$, which counts
only the crosses between the "sub-chains" of the length $\tau$
located near $i$ and $j$. It is defined as
\begin{equation}
  n_{ij}(x_{i\tau}-x_{j\tau},~x_{i\tau}-x_{(i+1)\tau},
~x_{j\tau}-x_{(j+1)\tau})=<\sum_{k,m}  \delta(x_k;x_m)>;
\end{equation}
where indices $k$ and $m$ run only within the intervals
$i\tau~...~i(\tau+1)-1$
and $j\tau~...~j(\tau+1)-1$, respectively, whereas the
averaging is carried out over the paths
passing through $x_{i\tau},x_{(i+1)\tau},x_{j\tau}$, and $x_{(j+1)\tau}$.

Full path $L$ can be divided into $L/\tau$ parts of the length $\tau$,
so that, for example, partition function $Z$ takes the form
\begin{equation} \label{E5}
Z=Z_0(\tau) \sum \exp
  \left(-\sum_{i=0}^{L/\tau} \frac{D ||x_{i \tau}-x_{(i+1)\tau}||^2}{2 \tau}
    -\frac{\beta}{2} \sum_{i,j=0}^{L/\tau} n_{ij}\right).
\end{equation}
Here the pre-exponential sum is taken over all possible sets
$\{x_0, x_\tau, ..., x_{L-\tau}\}$.

Obviously, $n(r, \rho_1, \rho_2)$ peaks near $r=0$ and falls fast at $r\to\infty$.
It can be verified \cite{ident} that $\sum_r n(r, \rho_1, \rho_2)=\tau^2$
at any $\rho_1, \rho_2$.
So the continuous limit of (\ref{E5}) is as follows:
\begin{equation} \label{cont}
  Z=Z_0 \int [{\cal D} x] \exp \left(-\int_0^L \frac{D x'^2(l)}{2} dl -\frac{\beta}{2} \int_0^L \int_0^L \delta
  (x(l)-x(l')) dl dl'\right).
\end{equation}
Here $\delta$ is the Dirac $\delta$-function.

Self-parallel shifts of the whole chain are irrelevant for the
statistical properties under study, so we can take the
path-integral only over the closed paths with the center of mass fixed at zero:
\begin{equation} \label{cond}
u=\int_0^L x dl=0;~~~x(0)=x(L).
\end{equation}

Continuous analog of the formula (\ref{avs}) for statistical averages  can be
written in the same manner. In particular, the distribution
function of $x$ is equal to
\begin{equation} \label{p}
p(x_0)=Z_p(x_0)/Z,
\end{equation}
where $Z_p(x_0)$ is a contribution to $Z$ from the trajectories
passing though the given point $x_0$ at $l=0$, that is $x(0)=x_0$.
Function $p(x_0)$ is normalised:
\begin{equation}\label{norm}
  \int p(x_0) dx_0=1.
\end{equation}

The key assumption of the present model is to approximate the second
term in the exponent by the interaction with harmonic potential:
\begin{equation}\label{repl}
-\frac{\beta}{2} \int_0^L \int_0^L \delta(x(l)-x(l')) dl dl' \approx
\int_0^L\left(U_0 + \frac{D
\Omega^2 x^2(l)}{2}\right) dl,
\end{equation}
where $\Omega$ is defined in a self-consistent way:
\begin{equation} \label{sc}
  D \Omega^2=-\beta L D^{-1} (\nabla^2 p)|_{x_0=0}
\end{equation}
(the value of $U_0$ may remain undefined, as
it drops out of the final formulae).

After the approximation (\ref{repl}) is done, the distribution
function can be found straightforwardly.
The problem resembles Feynman's path-integral expressions
for the statistics of a quantum particle \cite{Feynman}, with however
an additional condition for trajectories $\int x dl=0$.
Let us introduce a notation
\begin{equation}
  S_{x(l)}=\int_0^L\left(\frac{D x'^2}{2}-U_0 - \frac{D
\Omega^2 x^2}{2}\right) dl
\end{equation}
and define the "minimal-action"
trajectory $\bar{x}(l; x_0)$, which satisfy (\ref{cond}) and delivers
a minimum of $S$ at given $\bar{x}(0)=x_0$.
The conditional minimisation results in the problem of a classical
oscillator in the external field
\begin{equation}
  \frac{d^2 \bar{x}}{d l^2}+\Omega^2 \bar{x} - f=0,
\end{equation}
where the value of $f$ should be chosen to fulfil the condition
(\ref{cond}).
The following minimal value of $S$ is obtained:
\begin{eqnarray}\label{Smin}
  S_{min}(x_0)&=&-U_0 L+\frac{6 D \Lambda(\omega) x_0^2}{L}, \\ \nonumber
  \Lambda(\omega)&=&\frac{\omega^2}{3}\left(\frac{1}{1-\omega {\rm
  ctg}\omega}\right) \\ \nonumber
\omega&=&\Omega L /2.
\end{eqnarray}
Consider now possible deviations from this trajectory:
$x(l)=\bar{x}(l)+\tilde{x}(l)$, where $\tilde{x}(0)=0$ and
the conditions (\ref{cond}) are satisfied.
As the potential is harmonic it appears that
$S_x=S_{min}+S_{\tilde{x}}$, and $Z_p$ can be factorised:
\begin{equation}
  Z_p(x_0)=\tilde{Z} \exp(-S_{min}(x_0)).
\end{equation}
Here $\tilde{Z}$ does not depend on $x_0$. The last two formulae
and expressions (\ref{p}, \ref{norm}) allow to find out the
distribution function:
\begin{equation}
  p(x_0)=\left(\frac{6 D \Lambda(\omega)}{2 \pi L}\right)^{D/2}
       \exp\left(-\frac{6 D \Lambda(\omega) x_0^2}{L}\right)
\end{equation}
The dispersion is equal to
\begin{equation}\label {disp}
  <x^2>=\frac{L}{12 \Lambda(\omega)}.
\end{equation}
Substitution of $\nabla^2 p$ in formula (\ref{sc}) gives the equation
\begin{equation}\label{eq}
  \Lambda(\omega)=\frac{\pi}{3 D}
 \left(\frac{2 D \omega^2}{\pi \beta}\right) ^{\frac{2}{D+2}}
  L^{-\frac{4-D}{D+2}}.
\end{equation}

The left-hand and right-hand sides of this equation are
schematically plotted in Figure 2 as functions of $\omega$.
There is a single root within the range $(0,\pi)$. Its position is
governed by $\lambda=\beta^{2/(4-D)} L$. At $\lambda\ll 1$ we
obtain $\omega\approx0$ and therefore $\Lambda(\omega)\approx 1$,
so
\begin{equation} \label{as1}
  <x^2>|_{\lambda \ll 1}=\frac{L}{12}.
\end{equation}
The "uncorrelated" dependence $<x^2>\propto L$ is reproduced; moreover the
factor $1/12$ is also correct.
At the opposite limit $\omega(\lambda\gg1)\approx\pi$, and it follows from
(\ref{disp}-\ref{eq}) that
\begin{equation} \label{as2}
<x^2>|_{\lambda \gg 1} = \frac{D}{4 \pi}
\left(\frac{\beta}{2 \pi D}\right)^{\frac{2}{D+2}} L^{\frac{6}{D+2}}
\end{equation}
In particular $<x^2>\propto L$ in 4D and $<x^2>\propto L^2$ in 1D.
The indices in 2D and 3D are also quite reasonable.
In $D>4$ the scheme is hardly suitable, as
it is impossible to fulfill the inequality
$1 \ll \tau \ll \beta^{-2/(4-D)}$ and the
continuum analog (\ref{cont}) cannot be used.

It is important to discuss the following peculiarity of the
approximation made. Usually, the use of mean-field schemes means
that correlations like $<x(l_1) x(l_2)>$ at $l_1\neq l_2$ are totally neglected.
In our scheme however some correlations are still present,
because the center of mass of the trajectory is fixed (formula (\ref{cond})).
Namely that circumstance results in a non-trivial dependence of
$S_{min}(L)$ and the dimension-dependent critical index in the last
formula.

We have
also compared the result obtained with Monte-Carlo calculations in 1D and 2D.
Figure 3 shows bi-logarithmic plots
for the  dispersion $<x^2>$ {\it versus}
$L$ at $\beta=10^{-2}$ in 1D and $\beta=5 \cdot 10^{-2}$ in 2D.
Thin lines are asymptotes (\ref{as1},\ref{as2}). An
agreement is quite good even quantitatively.

The final remark is that although we have considered only
a particular random-walk
problem (\ref{E1}) with unitary steps, it is clear that
the model can be applied for any random-walk self-repulsing chains.
It is only required that (a) the repulsion is short-range and (b)
the central limit theorem can be applied,
yielding the Gaussian distribution (\ref{Gauss}).
Once these requirements are fulfilled, the continuous
analog of the universal form (\ref{cont}) can be established.

The author is grateful to P.A. Prudkovskii, A.A. Nikulin,
and A.A. Fedyanin, who made valuable stylistic comments.
The work was partially supported by
"Scientific Schools" program (grant 96-1596476).

\begin{figure}
\caption{Examples of the self-crossing closed chains in $D=2$. In the
central chain there is a single self-cross (positions of vertexes 1 and 3 are the same).
In the right chain there are 3 whose pairs (3-5, 3-7, and 5-7).}
\end{figure}

\begin{figure}
\caption{The sketch of $\Lambda(\Omega)$ (thick line) and
the right-hand side of the equation (19) (thin lines). Note that in $D<4$
the later is always a monotonic increasing function of $\omega$,
so there is always a single root of (19).
}
\end{figure}

\begin{figure}
\caption{Monte-Carlo calculations of the dispersion $<x^2>$ in 1D and 2D (points),
compared with estimations from formula (18). Thin lines show the displacive
and fractal asymptotes.}
\end{figure}


\begin{thebibliography}{99}
\bibitem {ident} One should place $\sum_r$ in the average brackets
and use the identity $\sum_x \delta(x,y)=1$.
\end{thebibliography}
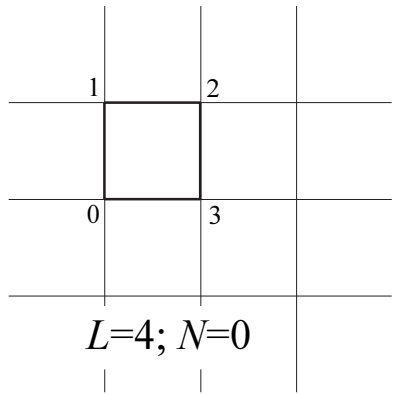
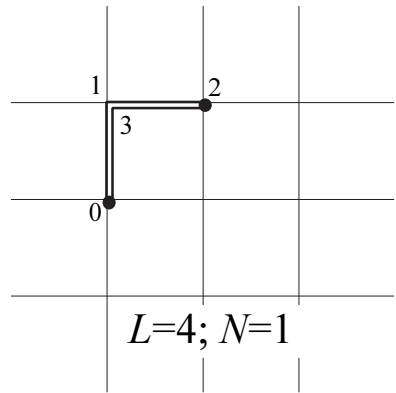
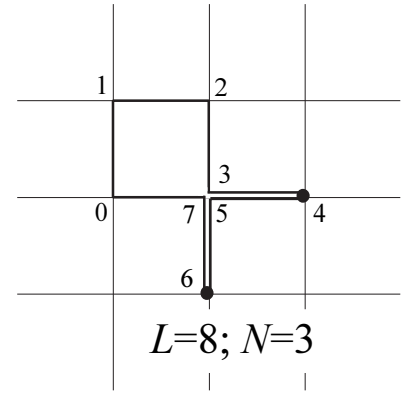
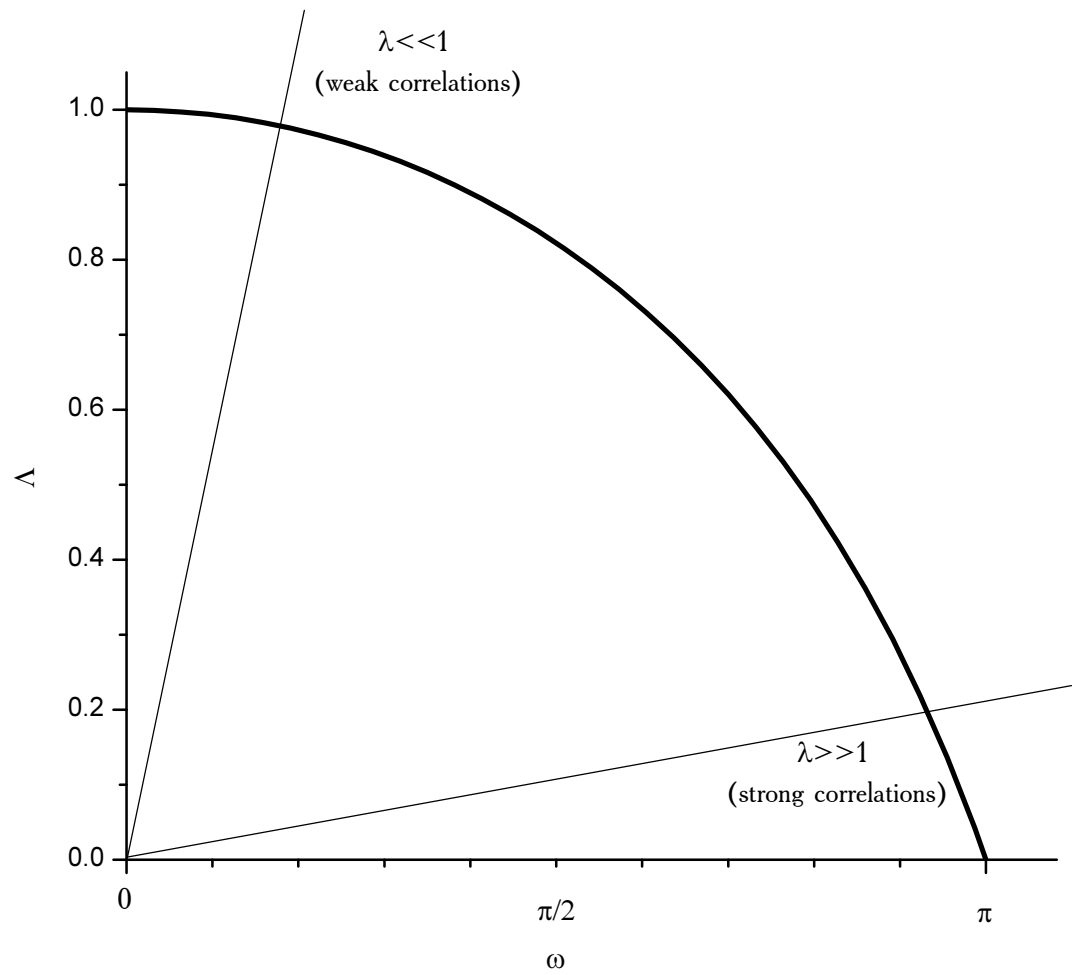
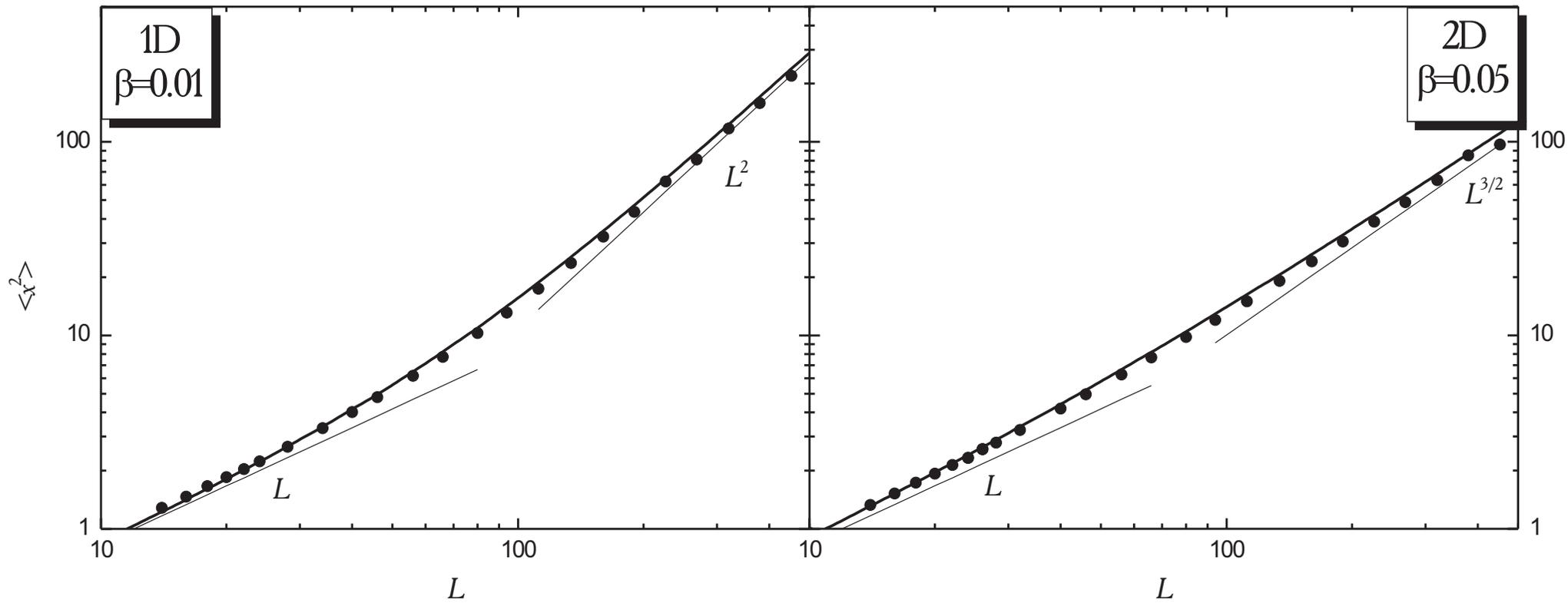
\end{document}